\documentstyle[11pt,aaspp4]{article}
\def\gsim{\;\rlap{\lower 2.5pt
 \hbox{$\sim$}}\raise 1.5pt\hbox{$>$}\;}
\def\lsim{\;\rlap{\lower 2.5pt
   \hbox{$\sim$}}\raise 1.5pt\hbox{$<$}\;}
\newcommand\beq{\begin{equation}}
\newcommand\eeq{\end{equation}}
\def\lya{Ly$\alpha$~}
\newcommand{\Ref}{\hangindent=20pt \hangafter=1 \noindent}
\newcommand{\StartRef}{\hyphenpenalty=10000 \raggedright
\parskip=0pt \parindent=0pt }

\begin{document}

\title{Determining the Redshift of Reionization From the Spectra of
High--Redshift Sources}

\author{Zolt\'{a}n Haiman\altaffilmark{1,2} and Abraham Loeb\altaffilmark{1}
\altaffiltext{1}{Astronomy Department, Harvard University, 60 Garden Street, Cambridge, MA 02138, USA}
\altaffiltext{2}{NASA/Fermilab Astrophysics Center, Fermi National Accelerator Laboratory, P.O. Box 500, Batavia, IL 60510, USA}}

\begin{abstract}
The redshift at which the universe was reionized is currently unknown.
We examine the optimal strategy for extracting this redshift, $z_{\rm
reion}$, from the spectra of early sources.  For a source located at a
redshift $z_{\rm s}$ beyond but close to reionization, $(1+z_{\rm
reion}) < (1+z_{\rm s}) < \frac{32}{27} (1+z_{\rm reion})$, the
Gunn--Peterson trough splits into disjoint Lyman $\alpha$, $\beta$,
and possibly higher Lyman series troughs, with some transmitted flux
in between these troughs. We show that although the transmitted flux
is suppressed considerably by the dense \lya forest after
reionization, it is still detectable for sufficiently bright sources
and can be used to infer the reionization redshift.  The Next
Generation Space Telescope will reach the spectroscopic sensitivity
required for the detection of such sources.

\end{abstract}


\centerline{Submitted to {\it The Astrophysical Journal}, July 1998}

\section{Introduction}

The standard Big Bang model predicts that the primeval plasma
recombined and became predominantly neutral as the universe cooled
below a temperature of several thousand degrees at a redshift
$z\sim10^3$ (Peebles 1968).  Indeed, the recent detection of Cosmic
Microwave Background (CMB) anisotropies rules out a fully ionized
intergalactic medium (IGM) beyond $z\sim300$ (Scott, Silk \& White
1995).  However, the lack of a Gunn--Peterson trough (GP, Gunn \&
Peterson 1965) in the spectra of high--redshift quasars (Schneider,
Schmidt \& Gunn 1991) and galaxies (Franx et al. 1997) implies that
the IGM is highly ionized at low redshifts, $z\lsim 5$.  Taken
together, these two observations indicate that the IGM was reionized
in the redshift interval $5\lsim z_{\rm reion} \lsim 300$.  The epoch
of reionization marks the end of the ``dark ages'' during which the
cosmic radiation background was dominated by the ever-fading CMB at
all wavelengths. During this epoch, ionized hydrogen (HII) started to
occupy again most of the volume of the universe.  Most likely, the
phase transition from HI to HII was triggered by the emission from the
first stars and quasars in the universe (see Loeb 1998, and references
therein).

The characteristic distance between the sources that ionized the
universe was much smaller than the Hubble scale, and the overlap of
the HII regions surrounding these sources was completed on a timescale
much shorter than the Hubble time.  The resulting reionization
redshift, $z_{\rm reion}$, is therefore sharply defined and serves as
an important milestone in the thermal history of the
universe. Measurement of this parameter would also constrain the
small-scale power of primordial density fluctuations that produced the
first objects.  Theoretical models and three--dimensional simulations
predict that reionization occurs at $7 \lsim z_{\rm reion} \lsim 20$,
just beyond the horizon of current observations (Haiman \& Loeb 1997,
1998; Gnedin \& Ostriker 1997).  Forthcoming instruments, such as the
{\it Space Infrared Telescope Facility} ({\it SIRTF}), and the {\it
Next Generation Space Telescope} ({\it NGST}) are expected to reach
the sensitivity required for the detection of sources beyond the
reionization redshift.  It is therefore timely to ask: how could one
infer $z_{\rm reion}$ from the observed spectra of these sources?

The spectrum of a source at a redshift $z_{\rm s}>z_{\rm reion}$
should show a GP trough due to absorption by the neutral IGM at
wavelengths shorter than the local \lya resonance at the source,
$\lambda_{\rm obs}<\lambda_\alpha(1+z_{\rm s})$. By itself, the
detection of such a trough would not uniquely establish the fact that
the source is located beyond $z_{\rm reion}$, since the lack of any
observed flux could be equally caused by (i) ionized regions with some
residual neutral fraction, (ii) individual damped \lya absorbers, or
(iii) line blanketing from lower column density \lya forest absorbers
(see, e.g. the spectrum of a $z=5.34$ galaxy taken by Dey et
al. 1998).  An alternative approach proposed by Miralda-Escud\'e
(1998) is to study the detailed shape of the damping wing of the GP
trough.  Although a measurement of this shape could ideally determine
the optical depth of the absorbing HI slab between the source redshift
and reionization, it is also likely to be compromised by contaminating
effects such as: (i) redshift distortions due to peculiar velocities;
(ii) uncertainties in the absorption profile due to the unknown
intrinsic shape of the \lya emission line from the source; (iii)
ionization of the IGM in the vicinity of the source due to the source
emission; and (iv) the existence of a damped \lya absorber outside or
inside the source.

In this paper we consider a different approach for measuring $z_{\rm
reion}$. Our method relies on the existence of some transmitted flux
between the Ly$\alpha$ and Ly$\beta$ absorption troughs in the spectra
of sources which are located just beyond the reionization redshift.
For such sources, the GP troughs due to different Lyman series
resonances do not necessarily overlap, and the transmitted flux
between these troughs is only affected by the post-reionization \lya
forest.  In \S 2, we simulate the absorption by the dense \lya forest
(or ``\lya jungle'') and demonstrate that the transmitted flux
features are detectable.  In \S 3, we estimate the number of sources
which are sufficiently bright to allow detection of these features
with {\it NGST}.  Finally, \S 4 summarizes the main conclusions of
this work.

\section{Spectra of High Redshift Sources}

The phase transition from HI to HII occurs almost simultaneously
throughout the IGM volume, when the expanding HII regions around the
separate ionizing sources overlap (Arons \& Wingert 1972).  This
process is nearly instantaneous (Haiman \& Loeb 1998; HL98) and
defines a corresponding redshift, $z_{\rm overlap}$.  However, at
$z_{\rm overlap}$ each individual HII region could still have a
non--negligible \lya optical depth due to its residual HI
fraction. The GP trough from the uniform IGM disappears only at a
somewhat later redshift, $z_{\rm reion}$, when the average \lya
optical depth of the smooth IGM drops below unity. (We distinguish
between this GP absorption and the absorption produced due to IGM
clumpiness by the post-reionization \lya forest).  In this paper, we
assume that the delay between $z_{\rm overlap}$ and $z_{\rm reion}$ is
small and set $z_{\rm overlap}=z_{\rm reion}$, based on the following
reasoning.

The \lya optical depth across a stationary HII sphere is as large as
$\sim10^4$ (Osterbrock 1974), but cosmological HII regions are much
thinner because of the cosmological redshifting of the \lya photons
away from resonance.  If $z_{\rm overlap}\sim7$ and the HII regions
are driven by quasars or star--clusters in Cold Dark Matter halos of
virial temperature of $\sim10^4$K ($M_{\rm
halo}\sim1.5\times10^{8}{\rm M_\odot}$), then the typical proper
radius of an HII region is $R_{\rm HII}\sim0.05$ Mpc (HL98).  When two
such HII regions first overlap, the neutral fraction $x=n_{\rm
HI}/n_{\rm H}$ at their intersection is $x\approx 2\times10^{-5}$,
assuming photoionization equilibrium\footnote{Note that the neutral
fraction would be $\sim$100 times higher near the edge of a
steady-state Str\"omgren sphere.  However, at redshifts $z\lsim10$ the
recombination time exceeds the Hubble time, and our short-lived
($\sim10^6$ years) sources do not reach a steady state.}.  Taking this
value as representative for the IGM, the resulting \lya optical depth
of the smooth IGM is $\tau_\alpha\approx3$.  The GP trough disappears
as soon as the neutral fraction is reduced by another factor of $\sim
3$, i.e.  when the local UV flux is increased by the same factor.
Such an increase occurs shortly after the overlap epoch, since it
requires the cumulative contribution from sources in a region as small
as $\sim 3\times 0.05~{\rm Mpc}=0.15$ Mpc, i.e.  $\sim0.1\%$ the local
Hubble length.  Therefore, the resulting breakthrough redshift,
$z_{\rm reion}$ is only slightly lower than $z_{\rm overlap}$.  The
short redshift delay distorts the shape of the damping wings at
wavelengths extending by a few tenths of a percent ($\lsim$ tens of
\AA) below the blue edge of the GP trough near the observed wavelength
of $\lambda_\alpha(1+z_{\rm reion})$.  Our simplifying assumption of
sudden ionization is therefore justified to this level of spectral
resolution.  Note that the Ly$\beta$ optical depth is $\sim$3.5 times
below that of Ly$\alpha$, so that the Ly$\beta$ and higher GP troughs
are even more sharply defined.

\begin{figure}[b]
\vspace{2.6cm}
\includegraphics{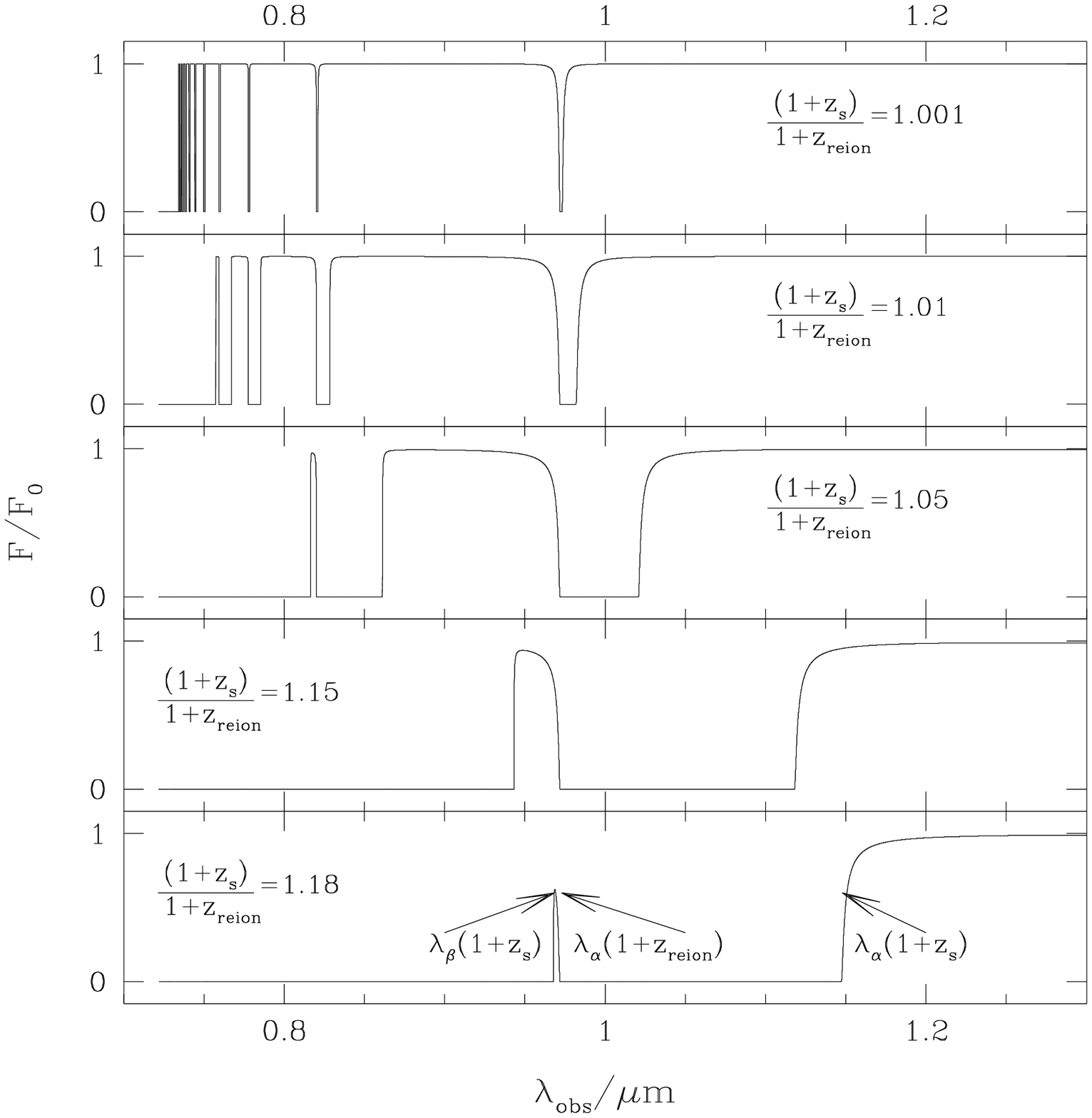}
\vspace*{4.5in}
\caption[Reionization Template] {\label{fig:template} Spectra of five
sources located beyond the reionization redshift (here taken to be $z_{\rm
reion}=7$), when all contaminating effects, such as peculiar velocities,
the ionization of the IGM in the vicinity of the source, or absorption by
the \lya forest, are ignored.}
\end{figure}

We now consider the spectrum of a source located at a redshift $z_{\rm
s}>z_{\rm reion}$.  \lya absorption by neutral hydrogen between
$z_{\rm reion}$ and $z_{\rm s}$ suppresses the flux in the observed
wavelength interval, $(1+z_{\rm reion})\lambda_\alpha < \lambda_{\rm
obs} < (1+z_{\rm s})\lambda_\alpha$; Ly$\beta$ absorption causes a
second trough at $(1+z_{\rm reion})\lambda_\beta < \lambda_{\rm obs} <
(1+z_{\rm s})\lambda_\beta$, and higher Lyman series lines produce
analogous troughs at shorter wavelengths.  In the absence of other
effects, the resulting spectrum would consist of a generic sequence of
troughs separated by blocks of transmitted flux, as depicted in
Figure~\ref{fig:template}.  In this illustration we assumed that
reionization occured suddenly as discussed above, at ${z_{\rm
reion}=7}$, but included the damping wings of the Lyman lines at both
edges of the absorption troughs (see the Appendix in Miralda-Escud\'e
1998).  Note that a Ly$\alpha$ photon emitted by the source can escape
from its host's HII region and travel across the uniform IGM only if
the HII region has an unusually large radius of at least $\sim$1 Mpc.
The emitted \lya photons are much more likely to be absorbed by the
IGM.  Below we consider observations on a telescope with a
sufficiently high angular resolution, so that the diffuse \lya
re-emission by the IGM can be ignored.

The top panel of Figure~\ref{fig:template} shows that the closer
$z_{\rm s}$ gets to $z_{\rm reion}$, the narrower the GP troughs
become and the more similar they look to individual \lya absorbers.
Equation~(\ref{eq:npress}) below predicts a few damped \lya systems
with neutral hydrogen column densities $N_{\rm H}\gsim 10^{20}~{\rm
cm^{-2}}$ in the spectra of a source at redshifts $z_{\rm s}\gsim 5$.
A single absorber with $N_{\rm H}=10^{20}~{\rm cm^{-2}}$ and
$b=35~{\rm km~s^{-1}}$ would have a rest--frame equivalent width of
$7.7$\AA.  In order to avoid confusion with such an absorber, the
width of the GP troughs must be $\gsim8(1+z_{\rm s})$\AA. Note also
that peculiar velocities could shift the location of the narrow
troughs by $\sim10$\AA. The bottom panel of Figure~\ref{fig:template}
also indicates that as the source redshift approaches the limiting
value $(1+z_{\rm s})/(1+z_{\rm
reion})=\lambda_\beta/\lambda_\alpha=32/27$, only a single narrow
block of transmitted flux remains between the \lya and Ly$\beta$
troughs. At still higher redshifts this window disappears, leaving a
continuous GP trough at wavelengths $\lambda_{\rm
obs}<\lambda_\alpha(1+z_{\rm reion})$ which carries no information
about the reionization redshift.

Reionization also leaves a distinct mark in the soft X-ray regime.
The source spectrum would be strongly suppressed at the Lyman edge,
$\lambda_{\rm obs}=912(1+z_{\rm s})$\AA, due to the large continuum
optical depth in from H and He ionizations ($\tau_{\rm
cont}\sim10^5$). The optical depth drops at shorter wavelengths due to
the sharp decline in the ionization cross--sections,
$\sigma\propto\lambda^3$. The flux recovers its intrinsic amplitude at
$E\gsim0.1$ keV, where the optical depth falls below unity.  The
observed photon energy where this occurs provides a measure of the
reionization redshift, as long as the value of $\Omega_{\rm b} h^2$ is
independently measured.  Would {\it AXAF} be able to detect this
spectral signature?  For a source whose observed flux at $E=0.1$ keV
is 100 nJy, the {\it AXAF} count rate in the 0.08-0.15 keV band would
be $10^{-3}$ per second\footnote{We consider detection by the High
Resolution Camera and the Low Energy Transmission Grating, based on
the proposal planning toolkit at
http://asc.harvard.edu/cgi-bin/prop\_toolkit.cgi.}, sufficiently large
to collect 100 photons in $10^5$ seconds.  The ability of {\it AXAF}
to determine $z_{\rm reion}$ relies on the existence of sufficiently
bright sources beyond the reionization redshift.  In the case where
these sources are early quasars, their required infrared flux is
$\gsim 10 \mu$Jy. As illustrated by Figure~\ref{fig:sources} below,
such sources are expected to be very rare at $z>5$.

The spectra shown in Figure~\ref{fig:template} are highly idealized in
that they take account only of absorption by the homogeneous HI gas
between $z_{\rm s}$ and $z_{\rm reion}$.  In reality, the blocks of
transmitted flux shown in Figure~\ref{fig:template} will be suppressed
due to absorption by the residual HI islands in the clumped IGM which
produce the \lya forest after reionization.  In order to simulate the
effect of the dense \lya forest (or ``\lya jungle''), we created mock
catalogs of absorption lines with statistical properties constrained
by observational data at $z<5$.  Press \& Rybicki (1993) have analyzed
in detail the statistics of \lya absorbers observed between redshifts
$1.5<z<4.3$, and derived a convenient form for the mean number
$n(N_{\rm H},b,z)dN_{\rm H}dbdz$ of clouds with HI column density
between $N_{\rm H}$ and $N_{\rm H}+ dN_{\rm H}$, Doppler velocity
parameter between $b$ and $b+db$, and redshift between $z$ and $z+dz$,
\beq 
n(N_{\rm H},b,z)=2.63\times10^{-14}(1+z)^\gamma
\left(\frac{N_{\rm H}}{10^{14}~{\rm cm^{-2}}}\right)^{-\beta}
p(b)~~~~{\rm cm^2~km^{-1}~s}, \label{eq:npress} 
\eeq with
$\gamma=2.46$, $\beta=1.43$, and 
\beq
p(b)\propto\exp\left[-\frac{(b-b_0)^2}{2b_*^2}\right]~~~~~(b>0)
\label{eq:bdist}
\eeq 
with $\int_0^{\infty}p(b)db=1$, and the best--fit values $b_0=32~{\rm
km~s^{-1}}$, $b_*=23~{\rm km~s^{-1}}$.  We used a Monte Carlo approach
to generate a catalog of absorbers along the line of sight to the
source, with redshifts, column densities, and $b$--parameters in the
ranges $0\leq z_{\rm i}\leq z_{\rm reion}$; $2\times10^{12}~{\rm
cm^{-2}}\leq N_{\rm H,i}\leq 10^{20}~{\rm cm^{-2}}$; and $b_{\rm
i}\geq 0$, for $1\leq i\leq N_{\rm abs}=\int n dz dN_{\rm H} db$. The
probability distribution of these parameters was chosen according to
the above equations.  Although these equation are based on data from
redshifts $1.5<z<4.3$, in the absence of additional data at higher
$z$, we extend their validity to all redshifts below $z_{\rm reion}$ .
A more rigorous extrapolation of this equation to high redshift would
be to use three-dimensional numerical simulations of the
high--redshift IGM (e.g., Hernquist~et~al.~1996).  However, even such
a calculation would be highly uncertain due to the unknown evolution
of the flux and spectrum of the UV background at $z>5$.

In computing the observed flux $F$ from the source at a wavelength
$\lambda_{\rm obs}$ we include absorption by the first nine Lyman
series lines from each of the $N_{\rm abs}(z_{\rm reion})$ absorbers
along the line of sight,
\beq
\label{eq:flux}
F(\lambda_{\rm obs})=F_0 \prod_{\rm n\in[1,9]} 
\prod_{i\in[1,N_{\rm abs}]} e^{-\tau_{\rm i,n} U(x_{\rm i,n},a_{\rm i,n})}.
\eeq
Here $F_0$ is the continuum flux that would have been observed in the
absence of \lya absorbers; $\tau_{\rm i,n}=(\pi e^2/m_ec) N_{\rm H,i}
\lambda_{\rm n}f_{\rm n}/ b_{\rm i}$ is the integrated optical depth
for the ${\rm n^{th}}$ Lyman line of the ${\rm i^{th}}$ absorber; and
$\lambda_{\rm n}$ and $f_{\rm n}$ are the rest wavelength and
oscillator strength of this line.  The profile of each line is given
by the normalized Voigt function $U(x,a)$, where $x_{\rm
i,n}\equiv(c/b_{\rm i})[\lambda_{\rm obs}/(1+z_{\rm i})-\lambda_{\rm
n}]/\lambda_{\rm n}$; $a_{\rm i,n}\equiv\Gamma_{\rm n}\lambda_{\rm
n}/4\pi b_{\rm i}$; and $\Gamma_{\rm n}$ is the natural decay rate of
the ${\rm n^{th}}$ Lyman line (see Press \& Rybicki 1993 for more
details).

\begin{figure}[b]
\vspace{2.6cm}
\includegraphics{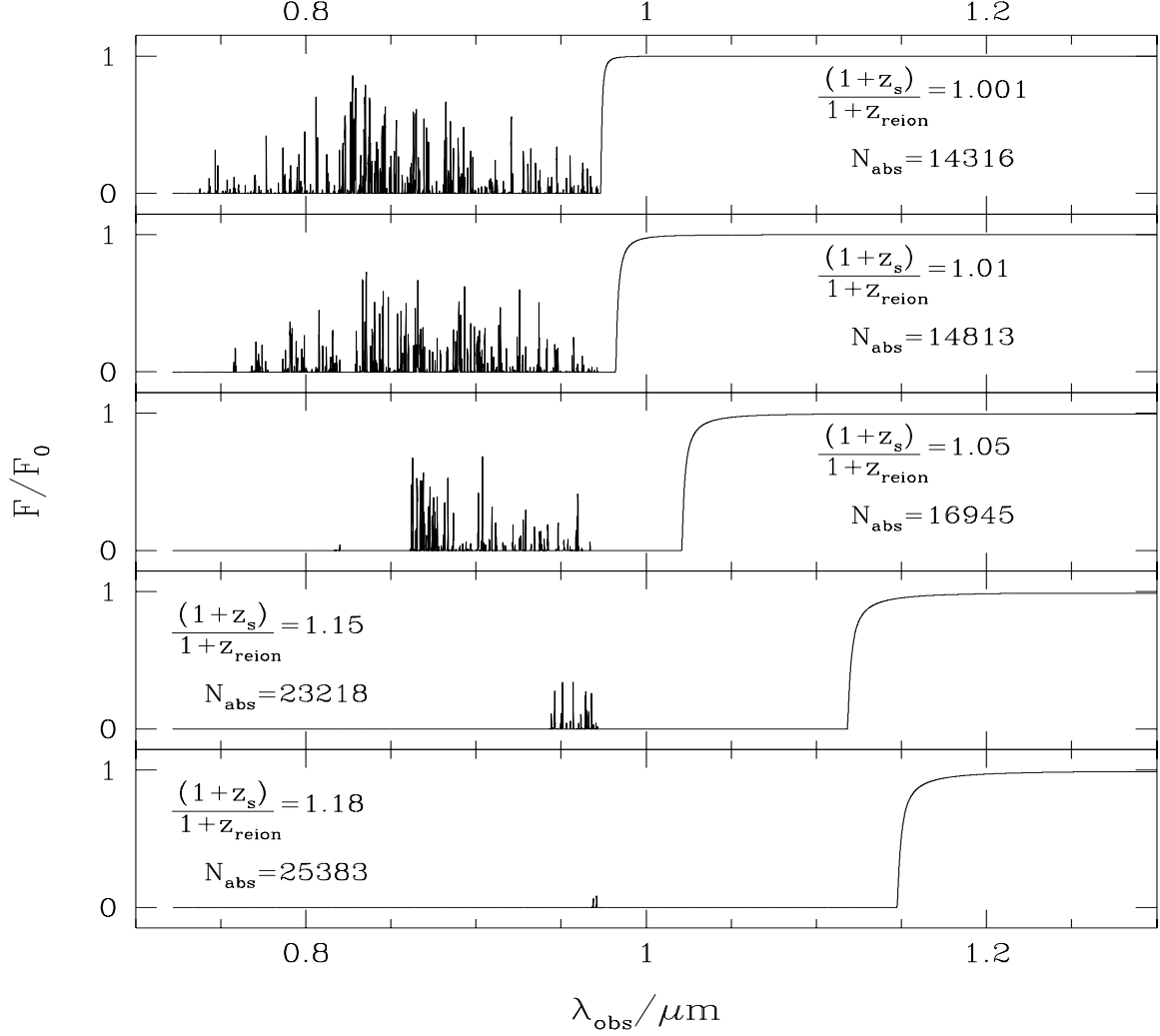}
\vspace*{4.5in}
\caption[Reionization Template Including Lyman Alpha Forest] 
{\label{fig:template2} Spectra of the 
five sources in Figure~\ref{fig:template}, 
when the absorption due to the post-reionization \lya forest is taken
into account.}
\end{figure}

Figure~\ref{fig:template2} shows five examples of source spectra
processed through the mock \lya jungle.  As in
Figure~\ref{fig:template}, we assume sudden reionization at $z_{\rm
reion}=7$ but include the damping wings of the GP troughs.  The
several thousand absorbers along a typical line of sight yield a
considerable mean optical depth, $\tau\approx 4.5$.  Although the
underlying continuum flux is not visible at any wavelength, the figure
reveals a series of narrow features of transmitted flux -- remnants of
the processing by the underlying \lya jungle -- to which we refer
hereafter as ``transmission features''.  The spectra contain many such
features with central flux values that rise infrequently up to a
significant fraction (up to $\sim50\%$) of the original continuum
flux.  These features could be detected by an instrument with a
sufficiently high sensitivity and spectral resolution.

\begin{figure}[b]
\vspace{2.6cm}
\includegraphics{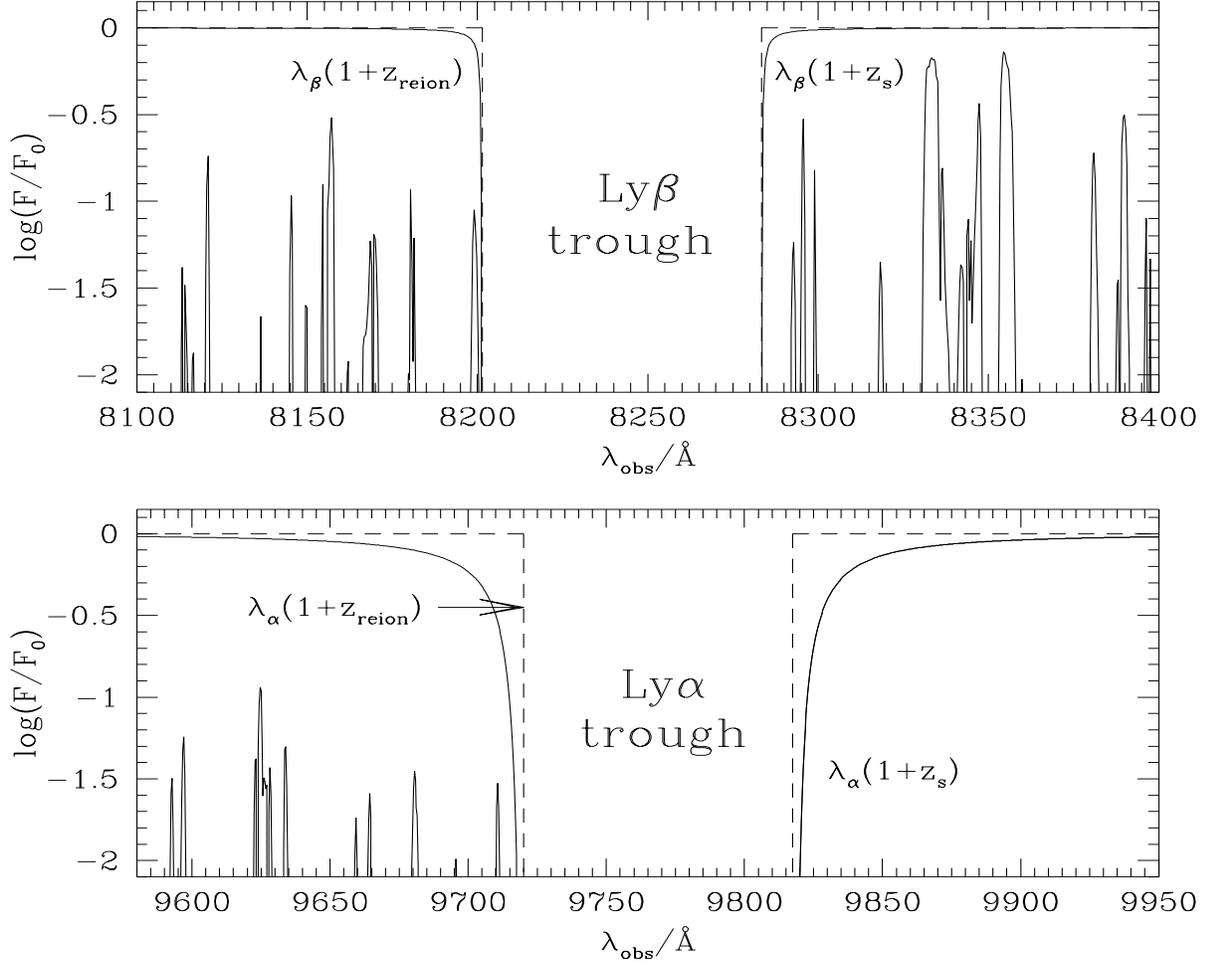}
\vspace*{4.5in}
\caption[Spectrum] {\label{fig:spect1} Blow--up of the spectrum of a
source at $z_{\rm s}=7.08$, assuming sudden reionization at a redshift
$z_{\rm reion}=7$.  The solid curves show the spectrum without
absorption by the high--redshift \lya forest, and the dashed lines
show the spectrum when the damping wings are also ignored.}
\end{figure}

As an example, we show in Figure~\ref{fig:spect1} a blow--up of the
spectrum around the Ly $\alpha$ and $\beta$ GP troughs of a source
with a redshift $z_{\rm s}=7.08$ .  The spectrum indeed contains
numerous transmission features; these features are typically a few
\AA~wide, have a central intensity of a few percent of the underlying
continuum, and are separated by $\sim 10$\AA.  With a sensitivity
reaching 1\% of the continuum, the edges of the Ly$\beta$ GP trough
between $8201$ and $8283$ \AA~could be identified to a $\sim10$
\AA~accuracy.  A similar precision applies to the Ly$\alpha$ GP
trough, although the damping wing there suppresses the flux within
$\sim10$ \AA~of the edges by a factor of $\sim3$.  The longest
wavelength at which flux is detected towards the blue edge of the
Ly$\beta$ GP trough, at $\sim8201$ \AA, provides the best direct
measurement of the reionization redshift.  In practice, this method
only yields a lower limit to $z_{\rm reion}$, since the first observed
transmission feature could be offset from the true blue edge of the GP
trough.  The uncertainty in the measurement of $\lambda_\beta(1+z_{\rm
reion})$ would be equal to the typical separation between the
transmission features.  Figure~\ref{fig:spect1} shows that at a
sensitivity reaching 1\% of the continuum, the separation would be
small ($\sim 10$\AA), leading to a fractional error of only
$\sim10/8200\sim0.1\%$ in the measurement of $z_{\rm reion}$.  Note
that this accuracy is similar to the level allowed by peculiar
velocities of a few hundred km/s, which could move either observed
edge of the GP trough by $\sim 10$\AA.

It is apparent from Figure~\ref{fig:spect1} that the above procedure
applies only as long as the typical separation between the
transmission features is much smaller than the width of the GP trough,
i.e. for $(1+z_{\rm s})\gsim 1.01\times(1+z_{\rm reion})$.  The bottom
three panels in Figure~\ref{fig:template2} also show that the
transmission features gradually disappear as the GP troughs of the
\lya and Ly$\beta$ resonances approach overlap at $(1+z_{\rm
s})/(1+z_{\rm reion})=\lambda_\alpha/ \lambda_\beta=32/27$.  Hence,
suitable sources must lie in the somewhat more restricted redshift
range, $1.01\lsim (1+z_{\rm s})/(1+z_{\rm reion})\lsim 1.17$.
Finally, we note that an extrapolation of equation~(\ref{eq:npress})
to even higher redshifts predicts that the mean optical depth of the
\lya clouds increases strongly with redshift.  For example, using our
mock \lya cloud catalog, we find that at wavelengths just below
$\lambda_\alpha(1+z_{\rm reion})$, the mean optical depth increases
from $\tau=4.5$ for $z_{\rm reion}=7$ to $\tau=13.1$ for $z_{\rm
reion}=10$.  Similar numbers are obtained by extrapolating the
expressions for the mean optical depth from Madau (1995).  This
implies that detection of the remnant transmission features is
considerably more difficult at higher redshifts, since the number
density of lines above a given sensitivity threshold drops as $\sim
\exp(-\tau)$.  In order to detect transmission features separated by
roughly $\sim 10$\AA, as in Figure~\ref{fig:spect1}, the spectroscopic
sensitivity must reach the approximate fractions of $\sim
10^{-1},10^{-2},10^{-3},10^{-4}$, and $10^{-6}$ of the continuum flux
for $z_{\rm reion}=6,7,8,9$, and $10$, respectively.

\section{Abundance of Suitable High Redshift Sources}

The detection of a single bright source from the epoch, $1.01
(1+z_{\rm reion}) < (1+z_{\rm s}) < 1.17 (1+z_{\rm reion})$, could in
principle provide an unambiguous measurement of $z_{\rm reion}$.  How
likely is it to find such sources?  The best candidates are
high--redshift quasars, since in analogy with their $z<5$
counterparts, they are expected to have a hard spectrum extending into
the far UV.  Alternative sources are Gamma-Ray Burst (GRB) afterglows,
supernovae, or primeval galaxies. The advantage of GRB afterglows is
that they are bright and have featureless power--law spectra; however,
their abundance at high redshifts might be small.  Supernovae and
galaxies may be more abundant than either quasars or GRBs, but are
likely to be faint and possess soft spectra in the relevant UV range.

As the abundance of quasars at $z>5$ is currently unknown, we must
resort to an extrapolation of the observed luminosity function (LF) of
quasars at $z\lsim5$.  Such an extrapolation has been carried out by
HL98 using a simple model based on the Press--Schechter (1974)
formalism. In the model of HL98, the evolution of the quasar LF at
faint luminosities and high redshifts is derived from three
assumptions: (i) the formation of dark--matter halos follows the
Press--Schechter theory, (ii) each dark halo forms a central black
hole with the same efficiency, and (iii) the light--curve of all
quasars, in Eddington units, is a universal function of time. As shown
in HL98, these assumptions provide an excellent fit to the observed
quasar LF at redshifts $2.6<z<4.5$ for an exponentially decaying
lightcurve with a time--constant of $\sim10^6$ years.  The model
provides a simple and natural extrapolation of the quasar LF to high
redshifts and faint luminosities.  At the faint end of the LF, the
number counts of quasars are expected to be reduced because of
feedback due to photoionization heating that prevents gas from
reaching the central regions of shallow potential wells. To be
consistent with the lack of faint point--sources in the Hubble Deep
Field, we imposed a lower limit of $\sim75~{\rm km~s^{-1}}$ for the
circular velocities of halos harboring central black holes (Haiman,
Madau \& Loeb 1998).

\begin{figure}[b]
\vspace{2.6cm}
\includegraphics{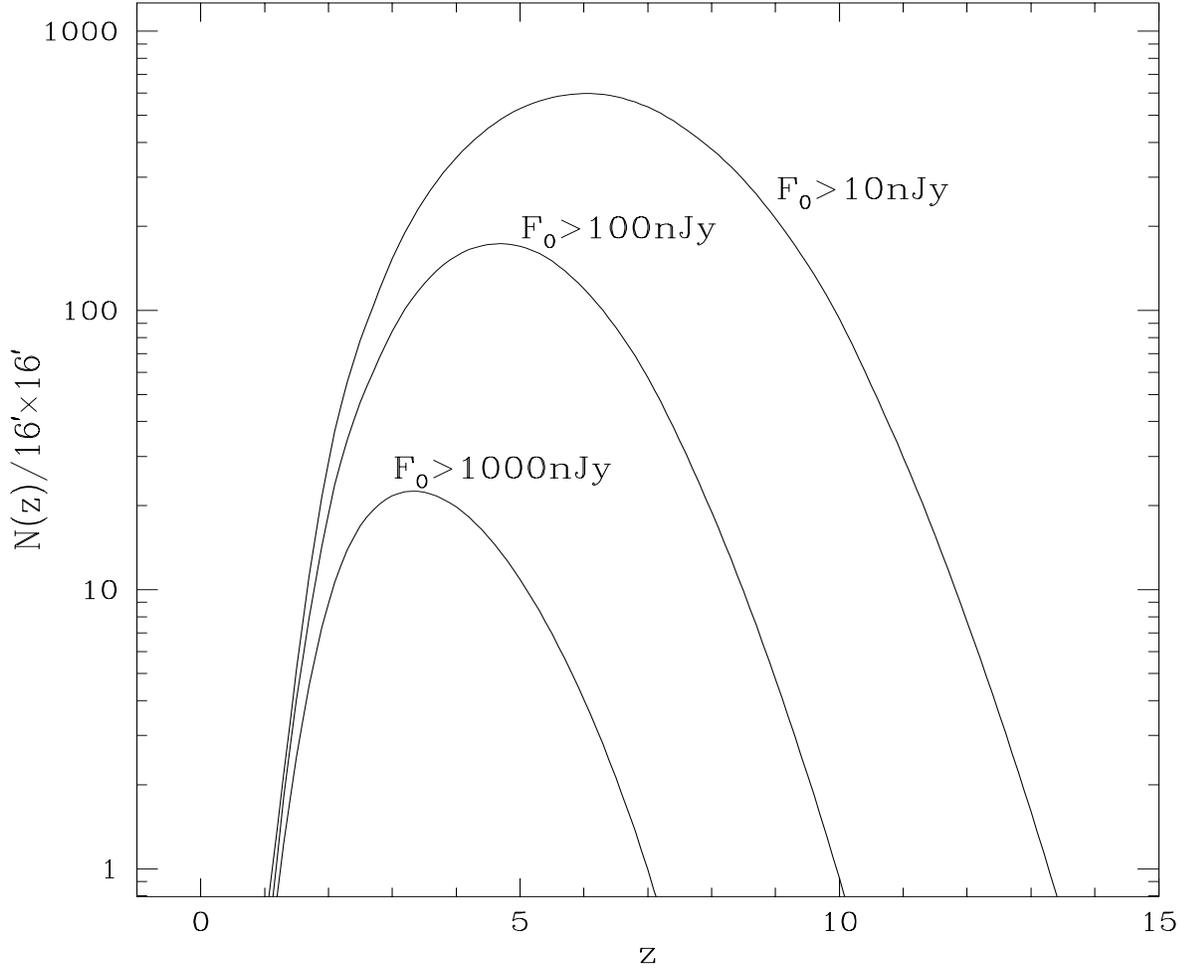}
\vspace*{4.5in}
\caption[Spectrum] {\label{fig:sources} Predicted number of quasars
within the redshift interval $1.01(1+z) < (1+z_{\rm s}) < 1.17 (1+z)$,
with a minimum intrinsic flux $F_0$, in a $16^\prime\times16^\prime$
field (sixteen $4^\prime\times4^\prime$ {\it NGST} fields of view).
The observed quasar luminosity function was extrapolated to high
redshifts using the model in HL98.}
\end{figure}

Figure~\ref{fig:sources} shows the total number $N(z,F)$ of quasars
brighter than the minimum flux $F$, located within the redshift
interval $1.01(1+z) < (1+z_{\rm s}) < 1.17(1+z)$, in a
$16^\prime\times16^\prime$ field (16 times the field of view of {\it
NGST}).  How many suitable sources would {\it NGST} detect?  For
$z_{\rm reion}=7$, the average transmitted flux shortward of the Lyman
$\alpha$ and $\beta$ GP troughs is a fraction $\exp(-\tau)\approx1\%$
of the continuum. Around $z\sim7$, we expect roughly a single
$\sim10^3$ nJy source per $16^\prime\times16^\prime$ field, based on
Figure~\ref{fig:sources}.  Such a source has an average flux of
$\sim$10 nJy blueward of $\lambda_\alpha(1+z_{\rm s}$ due to
absorption by the \lya jungle.  According to the method depicted on
Figure~\ref{fig:spect1}, a determination of $z_{\rm reion}$ to
$\sim1\%$ accuracy would require spectroscopy with a resolution of
$R$=100.  Requiring a signal-to-noise of S/N=3, the necessary
integration time for a 10 nJy source with {\it NGST} would be roughly
one hour\footnote{This result was obtained using the {\it NGST}
Exposure Time Calculator, available at http://augusta.stsci.edu. We
assumed a mirror diameter of 8m.}. This is a conservative estimate for
the detection of flux, since it assumes that the flux is uniformly
distributed across the spectrum, while in reality it would be
concentrated in transmission features that cover narrow wavelength
bins and yield a higher S/N ratio in these bins. We therefore conclude
that {\it NGST} will be able to measure the reionization redshift to
$\sim1\%$ percent accuracy up to $z_{\rm reion}=7$. The accuracy would
degrade considerably with redshift for $z_{\rm reion}\gsim7$.

The total number of sources per $16^\prime\times16^\prime$ field,
irrespective of their redshifts, is $\sim3\times 10^3$, $\sim600$, and
$\sim90$ for a minimum intrinsic flux of $F_{0}$=10, 100, and 1000
nJy, respectively.  In the most optimistic case, for which the
reionization redshift is close to the peak of the relevant curve in
Figure~\ref{fig:sources}, up to $\sim 25\%$ of all detected sources in
each image would lie in the redshift range $1.01(1+z) < (1+z_{\rm s})
< 1.17(1+z)$, and reveal the features depicted in
Figure~\ref{fig:template2}.  However, since only a small fraction of
all baryons need to be incorporated in collapsed objects in order to
ionize the IGM, reionization is likely to occur before the redshift at
which the source number-count peaks ($z_{\rm peak}=3.5, 4.5$, or $6$
at sensitivities of $10^3, 10^2$, or 10 nJy).  Pre--selection of
sources in particular redshift bins could be achieved photometrically
by bracketing the associated location of their \lya troughs with two
narrow-band filters.

\section{Conclusions}

We have shown that the Gunn--Peterson effect breaks into individual
Lyman series troughs in the spectra of sources located in the redshift
interval $1 < (1+z_{\rm s})/(1+z_{\rm reion}) < 32/27$. Although the
transmitted flux in between these troughs is heavily absorbed by the
Ly$\alpha$ jungle after reionization, the residual features it shows
allow a determination of the reionization redshift, $z_{\rm reion}$,
for sufficiently bright sources.  For a single source, a reionization
redshift $z_{\rm reion}\sim7$ could be determined to a precision of
$\sim1\%$ with a spectroscopic sensitivity of $\sim1\%$ of the
continuum flux.  A simple model for the abundance of high--redshift
quasars predicts that a single source per $16^\prime\times 16^\prime$
field could allow this detection at the $\sim10$ nJy spectroscopic
sensitivity of {\it NGST}.  It may also be feasible to probe
reionization with ground--based telescopes.  Using the Keck telescope,
Dey et al. (1998) have recently detected a source at $z$=5.34, and
found no continuum emission below the rest--frame \lya wavelength.
The authors were able to infer a lower limit of $\tau=1.2$ on the
optical depth of the \lya forest between 1050 and 1170 \AA~to
$z$=5.34, which falls just short of the $1.3<\tau<1.9$ implied by
equation~(\ref{eq:npress}) for this redshift and wavelength range.
Therefore, a somewhat more sensitive spectrum of this object could
potentially constrain reionization at $z<5.34$.

The signal-to-noise ratio in the determination of $z_{\rm reion}$
might be improved by co--adding the spectra of several sources.  In
the absence of \lya jungle absorption, the cumulative spectrum would
approach the sawtooth template spectrum derived by Haiman, Rees, \&
Loeb (1998, Fig.  1) and would be isotropic across the sky.  An
alternative signature of reionization should appear at the soft X-ray
region of the spectrum ($E\approx0.1$ keV), where the continuum
optical depth to H and He ionization drops below unity. {\it AXAF}
could detect such a signature in the spectra of only exceptionally
bright objects.

In order to enable a measurement of $z_{\rm reion}$ at redshifts as
low as 6, it is essential that the wavelength coverage of {\it NGST}
would extend down to $\sim7\lambda_\beta=0.7\mu$m.  Detection of the
transmitted flux features requires progressively higher sensitivities
as $z_{\rm reion}$ increases because of the likely increase in the
optical depth of the \lya jungle with redshift.  The nominal
integration time of about one hour for a $\sim10$ nJy sensitivity
(with $R$=100 and S/N=3) would be sufficient to determine $z_{\rm
reion}$ up to a redshift of $\sim7$ with a precision of $\sim1\%$.
Following the extrapolation of equation~(\ref{eq:npress}) to higher
redshifts, the required sensitivity needs to be one or two orders of
magnitude higher if $z_{\rm reion}\sim8$ or $\sim9$.  Note, however,
that if $z_{\rm reion}\gsim10$, then it will be more easily measurable
through the damping of the CMB anisotropies on small angular scales
(HL98). Near the planned spectroscopic sensitivity of {\it NGST}, most
sources are expected to have redshifts lower than $z_{\rm
reion}$. However, suitable high-redshift sources could be pre-selected
photometrically in analogy with the UV dropout technique
(Steidel~et~al~1996), by searching for the expected drop in the flux
at the blue edge ($\lambda_\alpha[1+z_{\rm reion}]$) of the \lya GP
trough.

\acknowledgements

This work was supported in part by the NASA ATP grants NAG5-3085,
NAG5-7039, and the Harvard Milton fund.  We thank George Rybicki for
advice on the statistics of the Ly$\alpha$ clouds, and Peter Stockman
for useful information about NGST.

\section*{REFERENCES}
{
\StartRef

\Ref Arons, J., \& Wingert, D. W. 1972, ApJ, 177, 1

\Ref Dey, A., Spinrad, H., Stern, D., Graham, J. R., \& Chaffee,
F. H. 1998, ApJL, in press, astro-ph/9803137

\Ref Franx, M. Illingworth, G. D., Kelson, D. D., Van Dokkum, P. G., 
\& Tran, K-V. 1997, ApJ, 486, L75
	
\Ref Gnedin, N. Y., \& Ostriker, J. P. 1997, ApJ, 486, 581

\Ref Gunn, J. E., \& Peterson, B. A., 1965, ApJ, 142, 1633

\Ref Haiman, Z., \& Loeb, A. 1997, ApJ, 483, 21 

\Ref Haiman, Z., \& Loeb, A. 1998, ApJ, in press, astro-ph/9710208 (HL98)

\Ref Haiman, Z., Madau, P., \& Loeb, A. 1998, ApJ, submitted, astro-ph/9805258

\Ref Haiman, Z., Rees, M. J., \& Loeb, A. 1997, ApJ, 476, 458 (HRL97)

\Ref Hernquist, L., Katz, N., Weinberg, D. H., \& Miralda-Escud\'e, J. 1996, ApJ, 457, L51

\Ref Loeb, A. 1998, in Science with the Next Generation Space Telescope, eds. 
E. Smith \& A. Koratkar, (ASP Conf. Series Publ., San-Francisco), p. 73;
preprint astro-ph/9704290

\Ref Madau, P. 1995, ApJ, 441, 18

\Ref Miralda-Escud\'e, J. 1998, ApJ, in press, preprint astro-ph/9708253

\Ref Osterbrock, D.E. 1974, Astrophysics of Gaseous Nebulae, (Freeman \& Co, 
San Francisco)
 
\Ref Peebles, P. J. E. 1968, ApJ, 153, 1

\Ref Press, W. H., \& Rybicki, G. B. 1993, ApJ, 418, 585

\Ref Press, W. H., \& Schechter, P. L. 1974, ApJ, 181, 425

\Ref Schneider, D. P., Schmidt, M., \& Gunn, J. E. 1991, AJ, 102, 837

\Ref Scott, D., Silk, J., \& White, M. 1995, Science, 268, 829

\Ref Steidel, C. C., Giavalisco, M., Dickinson, M., \& Adelberger, K. L. 1996, AJ, 112, 352

}


\clearpage
\newpage

\clearpage
\newpage

\clearpage
\newpage

\clearpage
\newpage
\end{document}